\documentclass[prd,twocolumn,showpacs,amsmath,amssymb,floatfix]{revtex4}
\usepackage{graphicx,color,dcolumn,booktabs,bm}
\usepackage{longtable,lscape}
\usepackage{txfonts}
\usepackage{amssymb}
\usepackage{indentfirst}

\begin{document}
\title{Nonresonant explanation for the $Y(4260)$ structure observed in the $e^+e^-\to J/\psi\pi^+\pi^-$ process}

\author{Dian-Yong Chen$^{1,2}$}
\author{Jun He$^{1,2}$}
\author{Xiang Liu$^{1,3}$\footnote{Corresponding author}}\email{xiangliu@lzu.edu.cn}
\affiliation{
$^1$Research Center for Hadron and CSR Physics,
Lanzhou University and Institute of Modern Physics of CAS, Lanzhou 730000, China\\
$^2$Institute of Modern Physics of CAS, Lanzhou 730000, China\\
$^3$School of Physical Science and Technology, Lanzhou University, Lanzhou 730000,  China}

\date{\today}
\begin{abstract}
In this work, we proposed a nonresonant explanation for the $Y(4260)$ structure observed in the $e^+e^-\to J/\psi\pi^+\pi^-$ process, {\it i.e.}, $Y(4260)$ is not a genuine resonance. Our result indicates that the $Y(4260)$ structure can be reproduced by the interference of production amplitudes of the $e^+e^-\to J/\psi\pi^+\pi^-$ processes via direct $e^+e^-$ annihilation and through intermediate charmonia $\psi(4160)/\psi(4415)$. Besides describing $Y(4260)$ line shape in $e^+e^-\to J/\psi\pi^+\pi^-$ well, such a nonresonant explanation for the $Y(4260)$ structure naturally answers why there is no evidence of $Y(4260)$ in the exclusive open-charm decay channel and $R$-value scan.
\end{abstract}

\maketitle

\section{Introduction}\label{sec1}

In the past seven years, different experimental collaborations have announced many
charmoniumlike states $X$, $Y$, $Z$. Among these observed
charmoniumlike states, $Y(4260)$ is the first structure reported in
the $e^+e^-\to J/\psi\pi^+\pi^-$ process \cite{Aubert:2005rm} at
BaBar. Later, both the CLEO Collaboration and the Belle Collaboration
confirmed $Y(4260)$ in the $e^+e^-\to J/\psi\pi^+\pi^-$ process
\cite{Coan:2006rv,He:2006kg,:2007sj}. Although CLEO found the first
evidence for $Y(4260)\to J/\psi K^+K^-$ \cite{Coan:2006rv}, Belle
indicated that no significant signal for $Y(4260)\to J/\psi K^+K^-$
was observed \cite{:2007bt}. Since the production of
$Y(4260)$ occurs via the $e^+e^-$ collisions with initial state
radiation, its spin-parity quantum number must be $J^{PC}=1^{--}$.
The mass and width of $Y(4260)$ from different experimental
measurements are listed in Table \ref{rs}.
\renewcommand{\arraystretch}{1.5}
\begin{table}[htb]
\caption{The measured mass and width of $Y(4260)$ by BaBar,
CLEO and Belle. \label{rs}} \label{table1}
\begin{center}
\begin{tabular}{cccccccccc}\toprule[1pt]
Experiment &Mass (MeV)& Width (MeV)\\\midrule[1pt]
BaBar \cite{Aubert:2005rm}& $4259\pm 8^{+2}_{-6}$ &$88\pm23^{+6}_{-4}$\\
CLEO \cite{He:2006kg}& $4284^{+17}_{-16}\pm 4$&$73^{+39}_{-25}\pm 5$\\
Belle \cite{:2007sj}& $4247\pm12^{+17}_{-32}$&$108\pm19\pm10$\\
Average \cite{PDG} & $4263^{+8}_{-9}$&$95\pm14$ \\ \bottomrule[1pt]
\end{tabular}
\end{center}
\end{table}

Since the $J/\psi\pi^+\pi^-$ invariant mass spectrum is not only
produced by the $e^+e^-$ annihilation but also by $B$ meson decay,
the BaBar Collaboration tried to find $Y(4260)$ by analyzing the
$J/\psi\pi^+\pi^-$ invariant mass spectrum from $B$ meson decay
\cite{Aubert:2005zh}. However, BaBar did not find any evidence for
$Y(4260)$ \cite{Aubert:2005zh}. Additionally, no evidence for
$Y(4260)$ was observed in the exclusive open-charm process
\cite{Abe:2006fj,Pakhlova:2008zza,Pakhlova:2007fq,:2009jv,Aubert:2006mi,:2009xs,CroninHennessy:2008yi}
and $R$-value scan from the BES Collaboration
\cite{Ablikim:2007gd}.

The observation of $Y(4260)$ has stimulated theorists' extensive
interest in understanding the underlying structure of $Y(4260)$.
Just because of the peculiar property of $Y(4260)$ revealed by
experiment, {\it i.e.}, $Y(4260)$ is observed  only in its hidden-charm
decay channel and is not found in the exclusive open-charm process, theorists
have proposed different exotic explanations for the structure of
$Y(4260)$, which include a charmonium hybrid
\cite{Zhu:2005hp,Kou:2005gt,Close:2005iz}, the first orbital
excitation of a diquark-antidiquark state ($[cs][\bar c\bar s]$)
\cite{Maiani:2005pe}, a $\chi_{c0}\rho^0$ molecule
\cite{Liu:2005ay}, an $\omega \chi_{c1}$ molecular state
\cite{Yuan:2005dr}, a $\Lambda_c\bar{\Lambda}_c$ baryonium state
\cite{Qiao:2005av}, the P-wave
$([cq]_{s=0}[\bar{c}\bar{q}]_{s=0})_{P-wave}$ tetraquark state
\cite{Ebert:2005nc,Ebert:2008kb}, a $D_1\bar{D}$ or $D_0\bar{D}^*$
molecular state \cite{Rosner:2006vc,Ding:2008gr}, a $J/\psi
K\bar{K}$ three-body system \cite{MartinezTorres:2009xb}, charmonium
hybrid state with strong coupling with $D\bar{D}_1$ and
$D^*\bar{D}_0$ \cite{Kalashnikova:2008qr}, and a $1S$ state of
a $D_1\bar{D}^*$ molecule \cite{Close:2009ag,Close:2010wq}. However, we note that
the lack of signal in certain channels also poses a serious challenge to a number of the explanations
proposed in the framework of an exotic state.

Besides explaining $Y(4260)$ as the exotic states just summarized
above, theorists have also tried to categorize $Y(4260)$ in the
charmonium family. In Ref. \cite{LlanesEstrada:2005hz}, the
possibility of $Y(4260)$ as $\psi(4260)$ corresponding to the
low member of the pair $4S-3D$ vector charmonium was discussed.
Eichten {\it et al.} indicated that $Y(4260)$ cannot be interpreted
as a conventional charmonium since the predicted total decay width
and the partial decay width into a charm meson pair are not consistent
with the experimental data if assuming $Y(4260)$ as the $2^3D_1$
$c\bar{c}$ state \cite{Eichten:2005ga}. Later, the result of the mass
spectrum in Ref. \cite{Segovia:2008zz} also shows that it is
difficult to put $Y(4260)$ into a conventional $c\bar{c}$ state. The
study of the mass spectrum of charmonium using a screened potential
indicates that the mass of $\psi(4S)$ is consistent with that of
$Y(4260)$ \cite{Li:2009zu}. At present, the main challenge of
$Y(4260)$ as a conventional charmonium is that we must answer why
there is no evidence of $Y(4260)$ in the obtained open-charm
process
\cite{Abe:2006fj,Pakhlova:2008zza,Pakhlova:2007fq,:2009jv,Aubert:2006mi,:2009xs,CroninHennessy:2008yi}
and $R$-value scan \cite{Ablikim:2007gd}, where the charmonia above 4 GeV should mainly decay into a charmed
meson pair.

Although theorists have discussed the structure of $Y(4260)$ under
the framework of exotic states or conventional charmonium,
recently a nonresonant explanation for $Y(4260)$ was suggested in
Refs. \cite{vanBeveren:2009fb,vanBeveren:2009jk,vanBeveren:2010mg},
where $Y(4260)$ is not a genuine resonance but rather a phenomenon
connected with the opening of the $D_s^*\bar{D}_s^*$ threshold and
the coupling to the $J/\psi f_0(980)$ and $J/\psi\sigma$ channels
\cite{vanBeveren:2009fb,vanBeveren:2009jk,vanBeveren:2010mg}.

At present, the key point of understanding the $Y(4260)$ structure
is that we must give a definite answer to explain why $Y(4260)$ is
only observed in its hidden-charm decay channel and is absent in the
observed open-charm decays. In addition, we should try our best to
reveal the properties of $Y(4260)$ by exhausting different
possibilities under the conventional framework.

In this work, we propose a nonresonant explanation for the
$Y(4260)$ structure observed in the $e^+e^-\to J/\psi\pi^+\pi^-$
process. In general, the $e^+e^-\to J/\psi\pi^+\pi^-$ process occurs via
two mechanisms. The first one is the direct production of $e^+e^-\to
J/\psi\pi^+\pi^-$ depicted in Fig.
\ref{Fig-Feyn1}(a), where $e^+e^-\to J/\psi\pi^+\pi^-$ directly occurs
without any intermediate charmonia. The second one is the
intermediate charmonium contribution to the $e^+e^-\to
J/\psi\pi^+\pi^-$ process, which is shown in Fig.
\ref{Fig-Feyn1}(b). Thus, we investigate whether the interference
effect between the amplitudes resulting from the above two mechanisms
is related to the $Y(4260)$ structure, which we will present in
detail in the next section.

This work is organized as follows. After the Introduction, we
illustrate the production of $e^+e^-\to J/\psi\pi^+\pi^-$ under two
mechanisms. In Sec. \ref{sec3}, the numerical results are given and compared with the experimental data. The last section is the
discussion and conclusion.

\section{The production amplitude for the $e^+e^-\to J/\psi\pi^+\pi^-$ process}\label{sec2}

\begin{figure}[htb]
\centering \scalebox{0.5}{\includegraphics{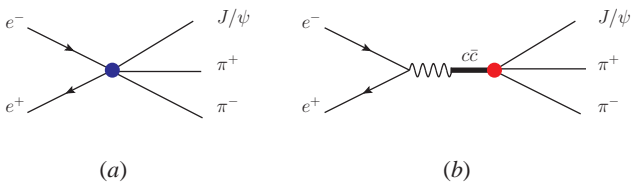}}
\put(-210,-20){$(a)$}%
\put(-80,-20){$(b)$}%
\caption{(color online). The diagrams relevant to $e^{+} e^{-} \to
J/\psi \pi^+ \pi^-$. Here, Fig. 1(a) corresponds to the $e^+e^-$
annihilation directly into $J/\psi\pi^+\pi^-$. Figure 1(b) is from the
contributions of intermediate charmonia.\label{Fig-Feyn1}}
\end{figure}

In this section, we illustrate how to obtain the production
amplitude of the $e^+e^-\to J/\psi\pi^+\pi^-$ process. As shown in
Fig. \ref{Fig-Feyn1}, there exists direct production of
$J/\psi\pi^+\pi^-$ by the $e^+e^-$ annihilation, which corresponds to
Fig. \ref{Fig-Feyn1}(a). The virtual photon from the $e^+e^-$ annihilation directly couples with
$J/\psi\pi^+\pi^-$.

For writing out the decay amplitude of Fig. \ref{Fig-Feyn1}(a), one constructs
the production amplitude to depict the direct production process of $e^+e^-\to J/\psi\pi^+\pi^-$
\begin{eqnarray}
\mathcal{M}_{{NoR}} = g_{{NoR}} \bar{u}(-k_1)
e\gamma_{\mu} u(k_2) \frac{1}{q^2}\epsilon_{J/\psi}^{\mu} (k_5)\,
\mathcal{F}_{{NoR}}(s),\label{1}
\end{eqnarray}
where the form factor $\mathcal{F}_{{NoR}}(s)$ \footnote{The form of
form factor $\mathcal{F}_{{NoR}}$ is not unique because we can also
have other choices, such as the dipole form factor
$\mathcal{F}_{{NoR}}(s)=(\sum_f m_f-b)^2/(\sqrt{s}-b)^2$, where $b$ is
the introduced parameter. Our study indicated that the experimental
data can also be described well in our model if adopting such dipole
form factor. (See the dashed line in
Fig. \ref{resulttotal} later in the paper for more details.) The fitting parameters are
$\beta_1=\beta_2=1$, $b=2.5587\pm0.0206$ GeV, $g_{_{NOR}}=0.2521\pm0.0553$~GeV,
$\phi_1=0.5519\pm0.5093$ Rad, $\phi_2=-1.2530\pm0.4713$ Rad,
$\phi_s=1.4488\pm0.4447$ Rad.} is introduced
to represent the $s$-dependence of $J/\psi \pi^+ \pi^-$ production directly via the
$e^+e^-$ annihilation, which can be represented as
$\mathcal{F}_{{NoR}} (s) =\mathrm{exp} \left(-a
(\sqrt{s}-\sum_f m_f)^2 \right)
$
with $\sum_f m_f$ as the sum of the masses of the final states for
$e^+e^-\to J/\psi\pi^+\pi^-$. In Eq. (\ref{1}), one introduces two
parameters $a$ and the coupling constant $g_{\mathrm{NoR}}$. $\sqrt{s}$
is the energy of the center of mass frame of $e^+e^-$. $k_1,\,k_2,\,k_5$
correspond to the four momenta of $e^+,\,e^-,\,J/\psi$,
respectively. $q=k_1+k_2$.

Besides the $e^+e^-$ annihilation directly into $J/\psi\pi^+\pi^-$,
another important production mechanism of $e^+e^-\to
J/\psi\pi^+\pi^-$ is through the intermediate charmonia, which is
shown in Fig. \ref{Fig-Feyn1}(b). As indicated by the
analysis of the $\pi^+\pi^-$ invariant mass spectrum of $e^+e^-\to
J/\psi\pi^+\pi^-$, $\pi^+\pi^-$ is from intermediate scalar states
$\sigma$ and $f_0(980)$ \cite{Aubert:2005rm}. Thus, the process of
$e^+e^-\to c\bar{c}\to J/\psi\pi^+\pi^-$ can be simplified as
$e^+e^-\to c\bar{c}\to J/\psi\mathcal{S}$, where $\mathcal{S}$
denotes the scalar states $\sigma$ and $f_0(980)$.

In the following, we need to depict the interaction between the
$c\bar{c}$ state and $J/\psi \mathcal{S}$ (solid [red] point shown in
Fig. \ref{Fig-Feyn1}(b), where the hadronic loop effect
can play a crucial role, which was proposed as an important
nonperturbative QCD mechanism to calculate the open-charm and
hidden-charm decays of charmonium \cite{Liu:2006dq,Liu:2009dr} or
a charmoniumlike state \cite{Liu:2006df,Liu:2008yy,Liu:2009iw}.

\begin{figure}[htb]
\centering
\begin{tabular}{cc}
\scalebox{0.6}{\includegraphics{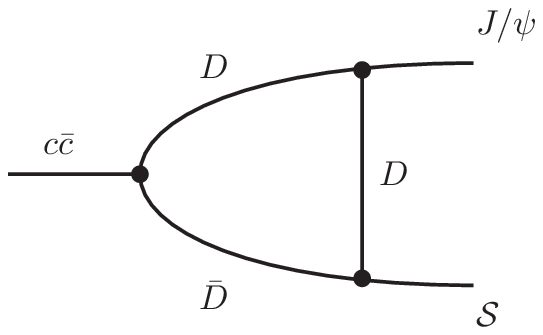}}&
\scalebox{0.6}{\includegraphics{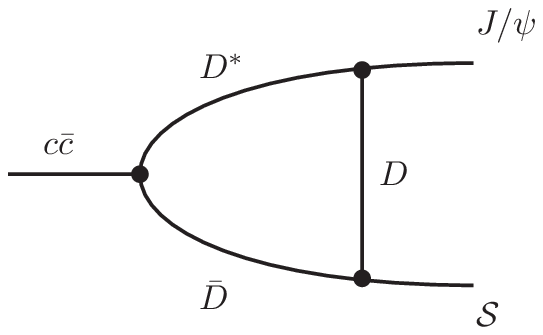}}\\
(a)&(b) \\
\scalebox{0.6}{\includegraphics{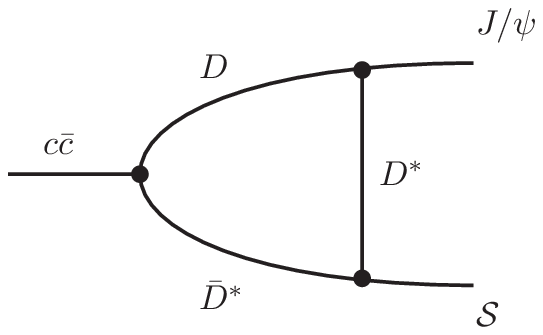}}&
\scalebox{0.6}{\includegraphics{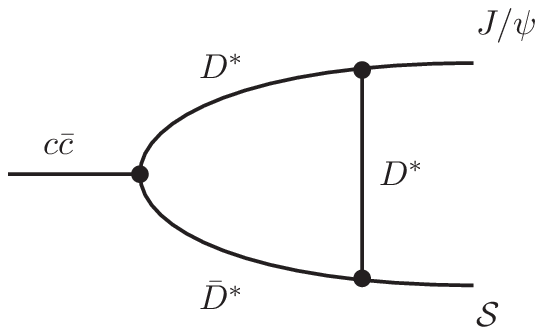}}\\
(c)&(d)
\end{tabular}
\caption{The typical diagrams relevant to charmonium interacting with
$J/\psi\mathcal{S}$ via the charmed meson loops.}
\label{Fig-Mesonloop}
\end{figure}

Under the hadronic loop mechanism, the detailed description of
charmonium coupling with $J/\psi\mathcal{S}$ is given in Fig.
\ref{Fig-Mesonloop}. The hadronic loops are constructed by charmed
mesons, which provide a bridge to connect the intermediate charmonium
and the $J/\psi\mathcal{S}$ final state.

For writing out the decay amplitudes corresponding to the diagrams
in Fig. \ref{Fig-Mesonloop}, we use the effective Lagrangian of vector charmonium interacting with
the charmed meson pair
\cite{Cheng:1992xi, Yan:1992gz, Wise:1992hn, Burdman:1992gh,
Casalbuoni:1996pg, Colangelo:2003sa}
\begin{eqnarray}
\mathcal{L}_{_{\psi \mathcal{D}\mathcal{D}}}&=&i g_{_{\psi
\mathcal{D}\mathcal{D}}} \psi_\mu \left(
\partial^\mu \mathcal{D} {\mathcal{D}}^{\dagger} - \mathcal{D}
\partial^\mu {\mathcal{D}}^{\dagger}
\right),\\
\mathcal{L}_{_{\psi \mathcal{D}^* \mathcal{D}}}&=&-g_{_{\psi
\mathcal{D}^* \mathcal{D}}}^{} \varepsilon^{\mu\nu\alpha\beta}
\partial_\mu \psi_\nu \left(
\partial_\alpha \mathcal{D}^*_\beta {\mathcal{D}}^{\dagger}
+ \mathcal{D} \partial_\alpha {\mathcal{D}}^{*\dagger}_\beta
\right),\\
\mathcal{L}_{_{\psi \mathcal{D}^* \mathcal{D}^*}}&=&-i g_{_{\psi
\mathcal{D}^* \mathcal{D}^*}}^{} \Bigl\{ \psi^\mu \left(
\partial_\mu \mathcal{D}^{*\nu} {\mathcal{D}}_\nu^{*\dagger} -
\mathcal{D}^{*\nu}
\partial_\mu {\mathcal{D}}_\nu^{*\dagger} \right)
\nonumber\\
&& + \left( \partial_\mu \psi_\nu \mathcal{D}^{*\nu}
-\psi_\nu
\partial_\mu \mathcal{D}^{*\nu} \right) {\mathcal{D}}^{*\mu\dagger}
\nonumber\\
&& + \mathcal{D}^{*\mu}\big( \psi^\nu
\partial_\mu {\mathcal{D}}^{*\dagger}_{\nu} - \partial_\mu \psi_\nu
{\mathcal{D}}^{*\nu\dagger} \big) \Bigr\},
\end{eqnarray}
and the Lagrangian of scalar state [$\sigma$ or $f_{0}(600)$] coupling with
$D^{(*)} D^{(*)}$
\begin{eqnarray}
\mathcal{L}_{_{SD^{(\ast)} D^{(\ast)}}} = g_{_{\mathcal{DD} \mathcal{S}}} \mathcal{S}
\mathcal{D} \mathcal{ D}^{\dag} - g_{_{\mathcal{D}^{\ast}
\mathcal{D}^{\ast} \mathcal{S}}} \mathcal{S} \mathcal{D}^{\ast} \cdot \mathcal{D}^{\ast
\dag},
\end{eqnarray}
where $\mathcal{D}=(\bar{D}^0,D^-,D_s^-)$ and $(\mathcal{D}^\dag)^T
= ({D}^0,D^+,D_s^+)$. The coupling constants involved in this work
will be presented in next section.

Thus, the obtained decay amplitudes for the diagrams in Fig.
\ref{Fig-Mesonloop} read as
\begin{eqnarray}
\mathcal{M}_{a} &=& (i)^3 \int \frac{d^4 q}{(2 \pi)^4} [i
g_{\psi DD} \epsilon_{\psi}^{\mu}(i p_{2 \mu}-i p_{1 \mu})]\nonumber\\&&\times
[i g_{J/\psi DD} \epsilon^{\rho}_{J/\psi} ( -i p_{1 \rho}- iq_{\rho})]
[g_{_{\mathcal{DDS}}}] \frac{1}{p_1^2-m_{D}^2} \nonumber\\
&&\times\frac{1}{p_2^2-m_{D}^2}
\frac{1}{q^2-m_{D}^2}\mathcal{F}^2(m_D^2,q^2)\mathcal{F}_{\psi}^{DD}(s),\label{eqa}
\end{eqnarray}
\begin{eqnarray}
\mathcal{M}_{b} &=&(i)^3 \int \frac{d^4 q}{(2 \pi)^4} [-g_{\psi
D^{\ast} D} \varepsilon_{\mu \nu \alpha \beta} (-i p_0^\mu)
\epsilon_{\psi}^\nu (i p_1^\alpha)] \nonumber\\&&\times[-g_{J/\psi D^\ast D}
\varepsilon_{\delta \tau \theta \phi} (i p_3^\delta)
\epsilon_{J/\psi}^\tau (- i p_1^\theta)] [g_{_{\mathcal{DDS}}}] \nonumber \\
&& \times \frac{-g^{\beta \phi}+p_1^\beta p_1^\phi
/m_{D^\ast}^2}{p_1^2 -m_{D^\ast}^2} \frac{1}{p_2^2 -m_{D}^2}
\frac{1}{q -m_{D}^2}\nonumber\\&&\times \mathcal{F}^2(m_D^2,q^2)\mathcal{F}_{\psi}^{D^*D}(s),
\end{eqnarray}
\begin{eqnarray}
\mathcal{M}_c &=& (i)^3 \int \frac{d^4 q}{(2 \pi)^4}  [- g_{\psi
DD} \varepsilon_{\mu \nu \alpha \beta} (-ip_0^\mu)
\epsilon_{\psi}^\nu (i p_2^\alpha)] \nonumber\\
&& \times[- g_{J/\psi DD}
\varepsilon_{\delta \tau \theta \phi} (ip_3^\delta)
\epsilon_{J/\psi}^\tau (i q^\theta) ] [-g_{_{\mathcal{D}^\ast \mathcal{D}^\ast \mathcal{S}}}]
 \frac{1}{p_1^2 -m_D^2} \nonumber\\
&& \times\frac{-g^{\beta \rho} + p_2^\beta
p_2^\rho /m_{D^\ast}^2}{p_2^2 -m_{D^\ast}^2} \frac{-g^{\phi \rho} +
q^\phi q^\rho /m_{D^\ast}^2}{q^2 -m_{D^\ast}^2} \nonumber\\&&\times\mathcal{F}^2(m_{D^*}^2,q^2)\mathcal{F}_{\psi}^{D^*D}(s),
\end{eqnarray}
\begin{eqnarray}
\mathcal{M}_{d} &=& (i)^3 \int \frac{d^4 q}{(2 \pi)^4} [-i g_{\psi
D^\ast D^\ast} \epsilon_{\psi}^{\mu} ((i p_{2 \mu}-i p_{1 \mu}
g_{\nu \rho})\nonumber\\
&&  + (-i p_{0 \rho}-i p_{2 \rho} g_{\mu \nu}) +(i p_{1
\nu}+i p_{0 \nu} g_{\mu \rho}) )] \nonumber \\
&& \times [-i g_{J/\psi D^\ast D^\ast} \epsilon_{J/\psi}^{\phi} ((-ip_{1
\phi}-iq_{\phi}) g_{\alpha \beta} +(ip_{3 \beta}+i p_{1 \beta})
g_{\alpha \phi}\nonumber\\
&&  +(iq_{\alpha}-i p_{3 \alpha}) g_{\beta \phi} )] [-
g_{_{\mathcal{D}^\ast \mathcal{D}^\ast \mathcal{S}}}]  \frac{-g^{\rho \alpha} +p_1^\rho p_1^\alpha /m_{D^\ast}^2}
{p_1^2 -m_{D^\ast}^2} \nonumber \\
&&\times \frac{-g^{\nu \tau} +p_2^\nu p_2^\tau
/m_{D^\ast}^2} {p_2^2 -m_{D^\ast}^2}  \frac{-g^{\beta \tau} +q^\beta
q^\tau /m_{D^\ast}^2} {q^2 -m_{D^\ast}^2}\nonumber\\&&\times\mathcal{F}^2 (m_{D^*}^2,q^2)\mathcal{F}_{\psi}^{D^*D^*}(s) ,\label{eqd}
\end{eqnarray}
where $\mathcal{F}(m_i^2,q^2)=(\Lambda_n^2-m_i^2)/(\Lambda_n^2-q^2)$
denotes the monopole form factor adopted in this work, which not
only compensates the off-shell effect of exchanged meson but also
describes the structure effect of the interaction vertex. Parameter
$\Lambda_n$ can be parametrized as $\Lambda_n= \beta_n
\Lambda_{QCD} +m_i$ with $\Lambda_{QCD}=220$ MeV and the mass $m_i$
of exchanged charmed meson, where $n$ in the subscript of $\beta_n$
is introduced to distinguish different $\beta$ values for different
charmonium transitions $c\bar{c}\to J/\psi\mathcal{S}$. Dimensionless parameter $\beta_n=1\sim 3$.
In addition, when we calculate the $e^+e^-\to c\bar{c}\to J/\psi \mathcal{S}$ process, we need to introduce another form factor to the interaction of the charmonium with $D^{(*)}D^{(*)}$ \cite{Pennington:2007xr}
\begin{eqnarray}
\mathcal{F}_{\psi}^{D^{(*)}D^{(*)}}(s) =\frac{\exp\Big[-\alpha_{\psi}^{D^{(*)}D^{(*)}} |\vec{p_1} (s,m^2_{D^{(*)}},m^2_{D^{(*)}})|^2\Big]
}{\exp\Big[-\alpha_{\psi}^{D^{(*)}D^{(*)}} |\vec{p_1} (m_\psi^2,m^2_{D^{(*)}},m^2_{D^{(*)}})|^2\Big]},
\label{Eq-FFs1}
\end{eqnarray}
which not only reflects the $|\vec{p_1}|$ dependence of the charmonium interacting with $D^{(*)}D^{(*)}$, but also represents the coupled channel effect summing up all the bubbles from the charmed meson loops \cite{vanBeveren:2007cb}. Here, $|\vec{p_1} (M^2,m^2_{D^{(*)}},m^2_{D^{(*)}})|=[\lambda(M^2,m^2_{D^{(*)}},m^2_{D^{(*)}})]^{1/2}/(2M)$ is the three-momentum of the intermediate charmed mesons in the center of the mass frame of intermediate charmonium with the K\"{a}llen function $\lambda(a,b,c)=a^2+b^2+c^2-2ab-2ac-2bc$. In the section on our numerical result, we will illustrate how to determine parameter $\alpha_{\psi}^{D^{(*)}D^{(*)}}$ in detail. $\mathcal{F}_{\psi}^{D^{(*)}D^{(*)}}(s)$ is normalized as 1 when $s=m_\psi^2$.

The total transition amplitude of $(c\bar{c}) \to J/\psi \mathcal{S}$
($\mathcal{S}=\sigma,\, f_{0}(980)$) is
\begin{eqnarray}
\mathcal{M}[(c\bar{c})\to J/\psi
\mathcal{S}]=2(\mathcal{M}_a+\mathcal{M}_b+\mathcal{M}_c+\mathcal{M}_d).\label{2}
\end{eqnarray}
Since the hadronic loops constructed by charge or neutral charmed
mesons can contribute to the $(c\bar{c})\to J/\psi \mathcal{S}$
transition, the factor 2 is introduced. Equation (\ref{2}) can be further simplified
as two independent Lorentz structures
\begin{eqnarray*}
\mathcal{M}[(c\bar{c})\to J/\psi \mathcal{S}]= \epsilon_{\psi}^{\mu}
\epsilon_{J/\psi}^{\nu} (g_A g_{\mu \nu} p_{_{J/\psi}} \cdot p_{\mathcal{S}} +
g_B p_{_{J/\psi} \mu} p_{\mathcal{S} \nu}),
\end{eqnarray*}
with two introduced coupling constants $g_{A}$ and $g_{B}$, which
are obtained by evaluating hadronic loop contributions in Eqs.
(\ref{eqa})-(\ref{eqd}). Here, $\epsilon_{\psi}$ and
$\epsilon_{J/\psi}$ are the polarization vectors of $(c\bar{c})$ and
$J/\psi$, respectively. $p_{\mathcal{S}}$ and $p_{_{J/\psi}}$ are
four momenta carried by scalar state and $J/\psi$, respectively. With the
above preparation and considering the vector meson dominance (VMD) mechanism \cite{Bauer:1977iq,Bauer:1975bw} for $\gamma\to (c\bar{c})$ coupling, we can write out the general amplitude of
$e^+(k_1)e^-(k_2)\to J/\psi(k_5)\pi^+(k_3)\pi^-(k_4)$ via the
intermediate charmonia
\begin{eqnarray}
\mathcal{M}_{\psi,\mathcal{S}} &=& \bar{u}(-k_1) e
\gamma^{\mu} u(k_2) \frac{-g_{\mu \nu}}{ (k_1 +k_2)^2} \frac{e\,
m_\psi^2/f_{\psi}}{(k_1+k_2)^2-m_{\psi}^2 + i
m_{\psi} \Gamma_{\psi}}  \nonumber\\
&&\times \epsilon_{J/\psi}^{\rho} \Big[g_{A} g_{\nu \rho} k_{5}
\cdot (k_3+k_4) +g_{B} k_{5 \nu} (k_{3 \rho} +k_{4 \rho}) \Big]
\nonumber\\
&&\times \frac{ g_{\mathcal{S} \pi \pi} }{(k_3+k_4)^2-m_{\mathcal{S}}^2 +
im_{\mathcal{S}} \Gamma_{\mathcal{S}}}(k_3
\cdot k_4), \label{dd}
\end{eqnarray}
where $f_{\psi}$ is the decay constant of intermediate charmonium
$c\bar{c}$. $g_{\mathcal{S} \pi\pi}$ denotes the coupling constant of
scalar state interacting with dipion.

We notice that the $Y(4260)$ structure is just sandwiched between
$\psi(4160)$ and $\psi(4415)$, which stimulates us to further
investigate whether the interference effect of direct production
amplitude $\mathcal{M}_{NoR}$ and the production amplitude
$\mathcal{M}_{\psi,\mathcal{S}}$ from the intermediate charmonia
$\psi(4160)$ and $\psi(4415)$ can reproduce the $Y(4260)$ structure
in $e^+e^-\to J/\psi \pi^+\pi^-$. Thus, in Eq. (\ref{dd}) we set
$\psi =\{\psi_1=\psi(4160), \psi_2=\psi(4415)\}$ and
$\mathcal{S}=\{\sigma, f_{0}(980)\}$.

The total amplitude for $e^{+} e^{-} \to J/\psi \pi^+ \pi^-$ is
described as
\begin{eqnarray}
\mathcal{M}_{tot} &=& \mathcal{M}_{NoR}+ e^{i \phi_1}
\Big(\mathcal{M} _{\psi_1, \sigma} + e^{i \phi_{s}} \mathcal{M}
_{\psi_1 f_0} \Big) \nonumber\\
&&\hspace{0mm}+ e^{i \phi_2} \Big(\mathcal{M} _{\psi_2, \sigma} +
e^{i \phi_{s}} \mathcal{M} _{\psi_2, f_0} \Big)\nonumber\\
&\equiv&\mathcal{M}_{NoR}+\mathcal{A}_{\psi_1}+\mathcal{A}_{\psi_2},\label{amp}
\end{eqnarray}
where we introduce phase angles $\phi_1$, $\phi_2$, $\phi_{s}$.
Generally speaking, the phases between different Feynman diagrams
are fixed and not arbitrary. However, there
maybe final state interactions which generate different phases among the
different diagrams because the momentum flow is different. Hence, it is
permissible to parametrize our ``ignorance" of these interactions with
an arbitrary phase to be fitted as is done in Eq. (\ref{amp}).
Observables are calculated by
summing the amplitudes and squaring.
Therefore, observables depend on
the sum of the squared amplitudes ($|\mathcal{M}_{NoR}|^2$, $|\mathcal{A}_{\psi_1}|^2$,
$|\mathcal{A}_{\psi_2}|^2$)
and cross terms ($2\mathrm{Re}(\mathcal{A}_{\psi_1}\mathcal{M}_{NoR}^*)$,
$2\mathrm{Re}(\mathcal{A}_{\psi_2}\mathcal{M}_{NoR}^*)$,
$2\mathrm{Re}(\mathcal{A}_{\psi_1}\mathcal{A}_{\psi_2}^*)$). The dependence of
the observables on the cross terms reflects the interference of production amplitudes of the $e^+e^-\to J/\psi\pi^+\pi^-$ processes via direct $e^+e^-$ annihilation and through intermediate charmonia $\psi(4160)/\psi(4415)$.

In the next section, we give the relevant numerical result.

\section{numerical result}\label{sec3}

The couplings of $J/\psi D^{(*)} D^{(*)}$ can be
obtained in the framework of heavy quark limit \cite{Oh:2000qr}.
Since $\psi(4160)/\psi(4415)$ is above the threshold of a pair of charmed
mesons, thus the coupling constants between $\psi(4160)/\psi(4415)$
and the charmed mesons can be evaluated by the partial decay width
of $\psi(4160)/\psi(4415) \to D^{(*)}D^{(*)}$ \cite{Barnes:2005pb}.
The determined coupling constants of $J/\psi$, $\psi(4160)$ and
$\psi(4415)$ coupling with $D^{(*)}D^{(*)}$ are listed in Table
\ref{Tab-cp}. The other coupling constants relevant to the
calculation include $g_{_{DD \sigma}}=g_{_{D^\ast D^\ast
\sigma}}=2m_D g_\sigma$ and $g_{_{DD f_0}}=g_{_{D^\ast D^\ast
f_0}}=2\sqrt{2}m_D g_\sigma$ with $g_\sigma=g_\pi/(2\sqrt{6})$ and
$g_\pi=3.73$ \cite{Bardeen:2003kt}. $g_{\sigma \pi \pi}= -0.02 \pm
0.002\ \mathrm{MeV} ^{-1}$ and $g_{f_0 \pi \pi}=-0.00354 \pm 0.0017\
\mathrm{MeV} ^{-1}$ are taken from Ref. \cite{Hooft:2008we}.

%
%
\begin{table}[htb]
\centering%
\caption{The coupling constants for $J/\psi$, $\psi(4160)$ and
$\psi(4415)$ coupling with $D^{(*)}D^{(*)}$ and the values of
$\alpha_\psi^{D^{(*)}D^{(*)}}$ in Eq. (\ref{Eq-FFs1}) for differnet
open-charm channels. The coupling constants of $\psi D D^{\ast}$ are
in unit of $\mathrm{GeV}^{-1}$.}
\label{Tab-cp}%
\begin{tabular}{cccc|cc}
\toprule[1pt]
     &\multicolumn{3}{c|}{Coupling constants}
     &\multicolumn{2}{c}{$\alpha_\psi^{D^{(*)}D^{(*)}}$ (GeV$^{-2}$)}\\
\cline{2-6}
     & $J/\psi$  & $\psi(4160)$ & $\psi(4415)$  & $\psi(4160)$ & $\psi(4415)$  \\
\midrule[1pt]
 $DD$            & $7.71$ & $1.86$  & $0.21$  & $0.70$ & $0.45$ \\
 $DD^\ast $      & $8.64$ & $0.10$  & $0.13$  & $1.20$ & $0.70$ \\
 $D^\ast D^\ast$ & $7.71$ & $1.50$  & $0.42$  & $2.20$ & $1.05$ \\
\bottomrule[1pt]
\end{tabular}
\end{table}

Besides the above coupling constants used in this work, we need to
determine the value of the parameter $\alpha_\psi^{D^{(*)}D^{(*)}}$ in
Eq. (\ref{Eq-FFs1}), which describes the effective radius of the
interaction between the charmonium and $D^{(\ast)} D^{(\ast)}$. Since
the form factor $\mathcal{F}_{\psi}^{D^{(*)}D^{(*)}}(s)$ in Eq.
(\ref{Eq-FFs1}) also plays a role to balance the over-increased
decay rates with increased phase space \cite{Meng:2008dd}, we choose
suitable values of parameter $\alpha_\psi^{D^{(*)}D^{(*)}}$ for
different open-charm channels (see Table \ref{Tab-cp}), which result
in the weak dependence of the corresponding open-charm decay ratios
of the charmonium on energy $\sqrt{s}$ in $e^+e^-\to c
\bar{c} \to J/\psi\pi^+\pi^-$ \cite{Meng:2008dd}.
%
%
\begin{table}[htb]
\centering%
\caption{The resonance parameters (in units of MeV) used in this
work. Here, the resonance parameters for charmonia relevant to this
work are taken from \cite{Ablikim:2007gd}. \label{Tab-Resonance}}
\begin{tabular}{cccccccc}
\toprule[1pt]
 $m_{\psi(4160)}$ & $4191.7$ & $\Gamma_{\psi(4160)}$ & $71.8$ &
 $m_{\sigma}$     & $526$    & $\Gamma_{\sigma}$     & $302$ \\
 $m_{\psi(4415)}$ & $4415.1$ & $\Gamma_{\psi(4415)}$ & $71.5$ &
 $m_{f_0(980)}$   & $980$    & $\Gamma_{f_0(980)}$   & $61$\\
\bottomrule[1pt]
\end{tabular}
\end{table}

In Table \ref{Tab-Resonance}, we also show the input resonance
parameters used in this work. In our model, we set seven free
parameters
\begin{eqnarray*}
\beta_1,\,\beta_2,\,\phi_1,\,\phi_2,\,
\phi_{s},\,g_{NoR},\,a, \label{fit}
\end{eqnarray*}
which are introduced in Sec. \ref{sec2}. The fitting results of such parameters are listed in Table \ref{chi}, where the errors and central values of the corresponding parameters are given. By checking the errors and the central values of the parameters, we notice that the obtained line shape for the $J/\psi\pi^+\pi^-$ invariant mass is not sensitive to small changes in the free parameters. Here, $\beta_1$ and $\beta_2$ are parameters in the form factor $\mathcal{F}(m_i^2,q^2)$
corresponding to $\psi(4160)$ and $\psi(4415)$.

\begin{figure}[htb]
\includegraphics[bb=25 340 560 780,scale=0.5]{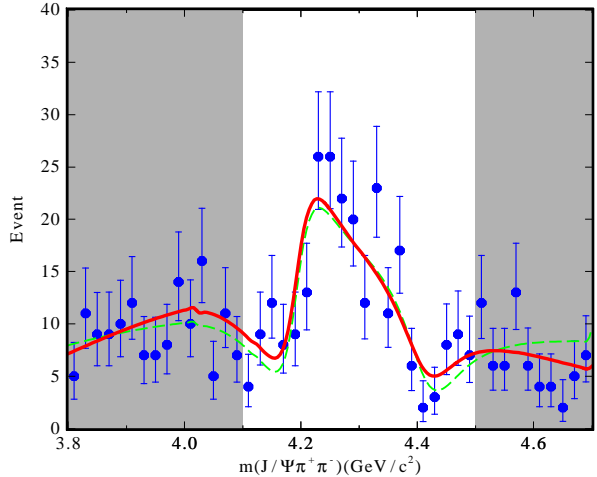}
\caption{(color online). The obtained fitting result (solid [red]
line) and the comparison with the experimental data (blue dots with
error bar) measured by BaBar \cite{Aubert:2005rm}. {We also give the obtained fitting result by adopting the dipole form for $\mathcal{F}_{NOR}(s)$ ([green] dashed line), where the expression of the dipole form factor and the values of the parameters are collected in [61].} Here, our result is normalized to the experimental data.
\label{resulttotal}}
\end{figure}

The main task of this work is to investigate whether we can
reproduce the $Y(4260)$ structure reported by the BaBar Collaboration in the
$e^+e^-\to J/\psi\pi^+\pi^-$ process \cite{Aubert:2005rm}. The
fitting result is presented in Fig. \ref{resulttotal} with the help
of the MINUIT package. Here, the binned maximum likelihood fit is
performed.  The obtained values of $\beta_i$ ($i=1,2$) fall just in
reasonable parameter range.

\begin{table}[htb]
\centering%
\caption{{The fitted parameters listed in Eq. (\ref{fit}).}
\label{chi}}
\begin{tabular}{crclcccc}
\toprule[1pt]
Parameter & Value (Rad)&Parameter & Value \\
\midrule[1pt]
$\phi_1$   &  $0.4545\pm0.3535$ & $g_{NoR}$  &   $0.0967\pm0.0280$ GeV  \\
$\phi_2$   &  $-0.9789\pm0.5146$ &  $a$        &   $0.7341\pm0.0678$ GeV$^{-2}$  \\
$\phi_3$   &  $1.5983\pm1.0922$ & $\beta_1,\,\beta_2$             &       1   \\
\bottomrule[1pt]
\end{tabular}
\end{table}

\begin{figure}[htb]
\includegraphics[bb=45 290 560 780,scale=0.50]{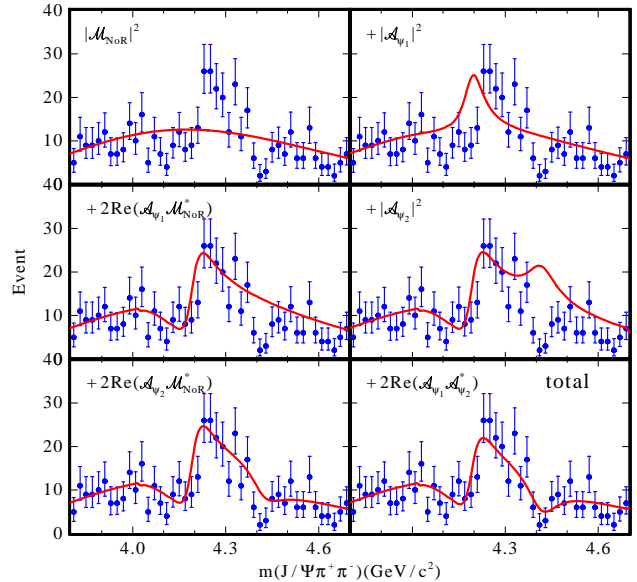}
\caption{(color online). The line shape of the cross section of the
$e^+e^-\to J/\psi\pi^+\pi^-$ process dependent on
$\sqrt{s}=m(\pi^+\pi^-J/\psi)$. Here, we present the changes of
the theoretical line shape by adding the contributions from these
six terms ($|\mathcal{M}_{NoR}|^2$, $|\mathcal{A}_{\psi_1}|^2$,
$2\mathrm{Re}(\mathcal{A}_{\psi_1}\mathcal{M}_{NoR}^*)$,
$|\mathcal{A}_{\psi_2}|^2$,
$2\mathrm{Re}(\mathcal{A}_{\psi_2}\mathcal{M}_{NoR}^*)$ and
$2\mathrm{Re}(\mathcal{A}_{\psi_1}\mathcal{A}_{\psi_2}^*)$) one by
one. The sum of these six terms finally results in the total line
shape of the cross section of the $e^+e^-\to J/\psi\pi^+\pi^-$ process
shown in the bottom right diagram.
 \label{result}}
\end{figure}

As shown in Fig. \ref{resulttotal}, the theoretical line shape
obtained by our model can fit experimental data well. In the
following, we present step by step how to get the theoretical line shape. Since $|\mathcal{M}_{tot}|^2$ in Eq. (\ref{amp}) can be
separated as six terms $|\mathcal{M}_{NoR}|^2$, $|\mathcal{A}_{\psi_1}|^2$,
$2\mathrm{Re}(\mathcal{A}_{\psi_1}\mathcal{M}_{NoR}^*)$,
$|\mathcal{A}_{\psi_2}|^2$,
$2\mathrm{Re}(\mathcal{A}_{\psi_2}\mathcal{M}_{NoR}^*)$ and
$2\mathrm{Re}(\mathcal{A}_{\psi_1}\mathcal{A}_{\psi_2}^*)$, the change
of the theoretical line shape is given in Fig. \ref{result} by
adding the contributions from these six terms one by one. Finally we
obtain the total line shape of the $e^+e^-\to J/\psi\pi^+\pi^-$ process.

The coupling constants relevant to the scalars $\sigma$ and $f_0(980)$ could not be same as those adopted in this work if one were using different approaches to describe the structure of scalars or a different calculation, such as dynamically generated resonance to explain scalar mesons \cite{Oller:1998zr,Gamermann:2006nm}. Our study indicates that such difference does not affect our fit result of the line shape for $J/\psi\pi^+\pi^-$ invariant mass, which can be compensated by the changes of free parameters set in our model.

\section{Conclusion and discussion}\label{sec4}

The experimental observation of charmoniumlike state $Y(4260)$
\cite{Aubert:2005rm} has stimulated extensive interest among
theorists and experimentalists. In the past seven years, different theoretical
explanations were proposed for understanding the underlying
structure of $Y(4260)$, which can be categorized in two groups,
{\it i.e.}, exotic state and the conventional charmonium explanations.

Although there already exist many theoretical explanations for
$Y(4260)$, we cannot give a definite solution to the $Y(4260)$
structure, which has spurred our interest in further investigating
$Y(4260)$ under a framework different from these existing theoretical
explanations.

The peculiar property of $Y(4260)$ is that $Y(4260)$ was only
observed in its hidden-charm decay channel $J/\psi\pi^+\pi^-$ by the
$e^+e^-$ annihilation. The present experimental measurements of
the $D^{(*)}\bar{D}^{(*)}$ invariant mass spectra and $R$-value scan do not show any evidence
of $Y(4260)$. Thus, if we want to solve $Y(4260)$, we must provide a
solution to explain why $Y(4260)$ has such peculiar properties.

Because the $Y(4260)$ structure is sandwiched between two known
charmonia $\psi(4160)$ and $\psi(4415)$, we proposed a nonresonant
explanation for the $Y(4260)$ structure. Here, $e^+e^-\to
J/\psi\pi^+\pi^-$ can occur directly via $e^+e^-$ annihilation. In
addition, the intermediate charmonia can contribute to $e^+e^-\to
J/\psi\pi^+\pi^-$, where we choose $\psi(4160)$ and $\psi(4415)$ as
intermediate charmonia. Thus, there exists an interference effect from
the above two production mechanisms for $e^+e^-\to
J/\psi\pi^+\pi^-$. In this work, we study whether such interference
effect can produce the line shape of $Y(4260)$ in the
$J/\psi\pi^+\pi^-$ invariant mass spectrum of the $e^+e^-\to J/\psi\pi^+\pi^-$ process.

The theoretical line shape and the comparison with experimental data
\cite{Aubert:2005rm} (see Figs. \ref{resulttotal}-\ref{result}) show
that the $Y(4260)$ structure can be reproduced by the interference
effect proposed in this work, where the intermediate charmonia
$\psi(4160)$ and $\psi(4415)$ play an important role in forming the
$Y(4260)$ structure. Thus, the nonresonant explanation for the
$Y(4260)$ structure proposed in this work is valuable to further
understanding the $Y(4260)$ structure. The nonresonant explanation to
the structure indicates that $Y(4260)$ is not a genuine
resonance, which naturally answers why experiment only reported
$Y(4260)$ in its hidden-charm decay channel.

In summary, at present the study of the charmoniumlike state $X$, $Y$,
$Z$ is an important research topic due to abundant experimental
information, which is full of challenges and opportunities. In this
work, with $Y(4260)$ as an example, we proposed a new approach
different from conventional charmonium and exotic state explanations
to explain $X$, $Y$, $Z$. Extending our model to further study
other charmoniumlike states $X$, $Y$, $Z$ is an interesting topic,
especially to these $X$, $Y$, $Z$ observed in the hidden-charm decay
channel from the $e^+e^-$ annihilation process \cite{Chen:2011kc}.

\section*{Acknowledgement}
We would like to thank two anonymous referees for many useful
suggestions and comments that have helped us clarify some points in the original version
of the paper. We also thank Professor Qiang Zhao for the discussion. This
project is supported by the National Natural Science Foundation of
China under Grants No. 10705001, No. 10905077, No. 11005129, No.
11035006, and No. 11047606 and the Ministry of Education of China
(FANEDD under Grant No. 200924, DPFIHE under Grant No.
20090211120029, NCET under Grant No. NCET-10-0442, and the Fundamental
Research Funds for the Central Universities).

\end{document}